\documentstyle[epsfig]{mn}

\newif\ifAMStwofonts
\AMStwofontstrue

\newcommand{\kms}{km\ s$^{-1}$}

\newcommand{\eg}{e.g.}
\newcommand{\cf}{cf.}

\newcommand{\etal}{et~al.}

\newcommand{\bJ}{b_J}
\newcommand{\mbJ}{m_{\bJ}}
\newcommand{\hcc}{Hydra-Centaurus Catalogue}
\newcommand{\ks}{Kolmogorov-Smirnov}
\newcommand{\GA}{Great Attractor}
\newcommand{\fog}{`Finger-of-God'}
\newcommand{\fogs}{`Fingers-of-God'}
\newcommand{\shap}{Shapley Supercluster}
\newcommand{\sgp}{Supergalactic Plane}

\newcommand{\nflair}{$4613$ }
\newcommand{\nz}{$3141$ }

\newcommand{\nearlyz}{$1272$ }

\newcommand{\nlatez}{$1802$ }

\newcommand{\somak}{Raychaudhury}

\ifoldfss
  \ifCUPmtlplainloaded \else
    \NewTextAlphabet{textbfit} {cmbxti10} {}
    \NewTextAlphabet{textbfss} {cmssbx10} {}
    \NewMathAlphabet{mathbfit} {cmbxti10} {} % for math mode
    \NewMathAlphabet{mathbfss} {cmssbx10} {} %  "   "    "
  \fi
  \ifAMStwofonts
    \ifCUPmtlplainloaded \else
      \NewSymbolFont{upmath} {eurm10}
      \NewSymbolFont{AMSa} {msam10}
      \NewMathSymbol{\upi}     {0}{upmath}{19}
      \NewMathSymbol{\umu}     {0}{upmath}{16}
      \NewMathSymbol{\upartial}{0}{upmath}{40}
      \NewMathSymbol{\leqslant}{3}{AMSa}{36}
      \NewMathSymbol{\geqslant}{3}{AMSa}{3E}

      \let\leq=\leqslant 
       \let\ge=\geqslant
    \fi
  \fi
\fi % End of OFSS

\ifnfssone
  \newmathalphabet{\mathit}
  \addtoversion{normal}{\mathit}{cmr}{m}{it}
  \addtoversion{bold}{\mathit}{cmr}{bx}{it}
  \newmathalphabet{\mathbfit} % math mode version of \textbfit{..}
  \addtoversion{normal}{\mathbfit}{cmr}{bx}{it}
  \addtoversion{bold}{\mathbfit}{cmr}{bx}{it}
  \newmathalphabet{\mathbfss} % math mode version of \textbfss{..}
  \addtoversion{normal}{\mathbfss}{cmss}{bx}{n}
  \addtoversion{bold}{\mathbfss}{cmss}{bx}{n}
  \ifAMStwofonts
    \ifCUPmtlplainloaded \else
      \UseAMStwoboldmath
      \makeatletter
      \new@mathgroup\upmath@group
      \define@mathgroup\mv@normal\upmath@group{eur}{m}{n}
      \define@mathgroup\mv@bold\upmath@group{eur}{b}{n}
      \edef\UPM{\hexnumber\upmath@group}
      \new@mathgroup\amsa@group
      \define@mathgroup\mv@normal\amsa@group{msa}{m}{n}
      \define@mathgroup\mv@bold\amsa@group{msa}{m}{n}
      \edef\AMSa{\hexnumber\amsa@group}
      \makeatother
      \mathchardef\upi="0\UPM19
      \mathchardef\umu="0\UPM16
      \mathchardef\upartial="0\UPM40
      \mathchardef\leqslant="3\AMSa36
      \mathchardef\geqslant="3\AMSa3E

      \let\leq=\leqslant 
       \let\ge=\geqslant
    \fi
  \fi
\fi % End of NFSS release 1

\ifnfsstwo
  \DeclareMathAlphabet{\mathbfit}{OT1}{cmr}{bx}{it}
  \SetMathAlphabet\mathbfit{bold}{OT1}{cmr}{bx}{it}
  \DeclareMathAlphabet{\mathbfss}{OT1}{cmss}{bx}{n}
  \SetMathAlphabet\mathbfss{bold}{OT1}{cmss}{bx}{n}
  \ifAMStwofonts
    \ifCUPmtlplainloaded \else
      \DeclareSymbolFont{UPM}{U}{eur}{m}{n}
      \SetSymbolFont{UPM}{bold}{U}{eur}{b}{n}
      \DeclareSymbolFont{AMSa}{U}{msa}{m}{n}
      \DeclareMathSymbol{\upi}{0}{UPM}{"19}
      \DeclareMathSymbol{\umu}{0}{UPM}{"16}
      \DeclareMathSymbol{\upartial}{0}{UPM}{"40}
      \DeclareMathSymbol{\leqslant}{3}{AMSa}{"36}
      \DeclareMathSymbol{\geqslant}{3}{AMSa}{"3E}

      \let\leq=\leqslant 
       \let\ge=\geqslant
    \fi
  \fi
\fi % End of NFSS release 2

\ifCUPmtlplainloaded \else
  \ifAMStwofonts \else % If no AMS fonts
    \def\upi{\pi}
    \def\umu{\mu}
    \def\upartial{\partial}
  \fi
\fi

\title{FLASH redshift survey---I. Observations and Catalogue}
\author[Raven Kaldare \etal]
       {Raven Kaldare$^1$, Matthew Colless$^2$, Somak \somak$^{3}$, B.A.\ Peterson$^2$\\
	$^1$ Institute of Astronomy, University of Cambridge, Madingley Road, Cambridge CB3 0HA, United Kingdom \\
	$^2$ Research School of Astronomy \& Astrophysics, The Australian National University, Weston Creek, ACT 2611, Australia \\
	$^3$ School of Physics \& Astronomy, University of Birmingham, Edgbaston, Birmingham B15 2TT, United Kingdom}

\date{Accepted ---.
      Received ---;
      in original form ---}

\pagerange{\pageref{firstpage}--\pageref{lastpage}}
\pubyear{2001}

\begin{document}

\maketitle

\label{firstpage}

\begin{abstract}
The FLAIR Shapley-Hydra (FLASH) redshift survey catalogue consists of
\nflair galaxies brighter than $\bJ = 16.7$ (corrected for Galactic
extinction) over a 605 sq. degree region of sky in the general
direction of the Local Group motion. The survey region is
approximately $60\degr
\times 10\degr$ strip spanning the sky from the \shap\ to the Hydra
cluster, and contains \nz galaxies with measured redshifts.  
Designed to explore   
the effect of the galaxy concentrations in this direction
(in particular the Supergalactic plane and the Shapley Supercluster)
upon the Local Group motion, the 68\%
completeness allows us to sample the large-scale structure better than 
similar sparsely-sampled surveys.
The survey region does not overlap with the areas covered by 
ongoing wide-angle (Sloan or
2dF) complete redshift surveys.  In this paper, the first in a series, we
describe the observation and data reduction procedures, the analysis
for the redshift errors and survey completeness, and present the
survey data.

\end{abstract}

\begin{keywords}
catalogues -- galaxies:distances and redshifts --
cosmology:observations -- galaxies:general --
-- large-scale structure of Universe
\end{keywords}

\section{Introduction}

This paper describes the FLASH redshift survey and presents the survey
catalogue.  The acronym FLASH is derived from the name of the primary
observing tool -- the FLAIR fibre spectrograph on the UK Schmidt
Telescope (UKST) -- and the names of the two concentrations of
galaxies, the \shap\ and the Hydra cluster, that lie at either end of
the survey region. As well as the structures associated with the
Shapley and Hydra-Centaurus superclusters, the survey region is
densely populated with other galaxy clusters, and lies across the
\sgp\ at its densest concentration.  It lies approximately in the
direction of motion of the Local Group with respect to the cosmic
microwave background, $(l,b)$=(236\degr,
30\degr)\cite{kogut:1993}. The core of the \GA\ itself
\cite{burstein:90,7s} lies just outside the survey region, at
$(l,b)$=(309\degr,18\degr).

The main goal of the FLASH survey is to investigate the detailed structure
of this densely-populated region and to infer the effect of the various
galaxy concentrations it contains on the motion of the Local Group (LG).
Indeed, the largest overdensity in the survey region, the \shap, is also
the largest overdensity within the local ($z\!<\! 0.1$) universe 
\cite{somak:1991,fabian:1991}, and lies in the general direction of
the LG motion. These properties of the supercluster have naturally lead to
speculations concerning its contribution to the motion of the LG.

Melnick \& Moles \shortcite{melnick:1987} were the first to examine the
effects of (what is now known to be) the \shap\ upon the LG's motion.
Using a simple analysis of the dynamics of the galaxies within the
supercluster, they concluded that the mass of the supercluster was not
sufficient to be the source of the LG's motion.  Analysis of the \shap's
contribution to the optical dipole \cite{somak:1989} revealed that the
supercluster's contribution to the LG motion was no more than 10 per
cent, while Quintana et al.  \shortcite{quintana:1995} found, by
estimating the supercluster's mass from the mass of the individual
clusters within, that the \shap\ may contribute up to 25 per cent of the
LG's motion.  Later estimates of the mass of the \shap\ from x-ray
clusters \cite{efw:1997} or from the number overdensity within its core
region by Bardelli et al.  \shortcite{bzzms:2000} have produced results
consistent with the above. However, from a recent redshift survey of the
\shap\ region, Drinkwater et al. \shortcite{drinkwater:1999} have
speculated that the \shap\ could be at least 50 per cent more massive
than earlier estimates.

Secondary goals of the survey include an examination of the variations
of the galaxy luminosity and correlation functions with morphological
type, taking advantage of the visual morphological classifications in
the parent sample, the \hcc\ \cite{somak:2002}.  Since our
sample spans a large range of local galaxy density, we will also
compare various properties of galaxies as a function of local
environment.

\begin{figure*}
\epsfig{file=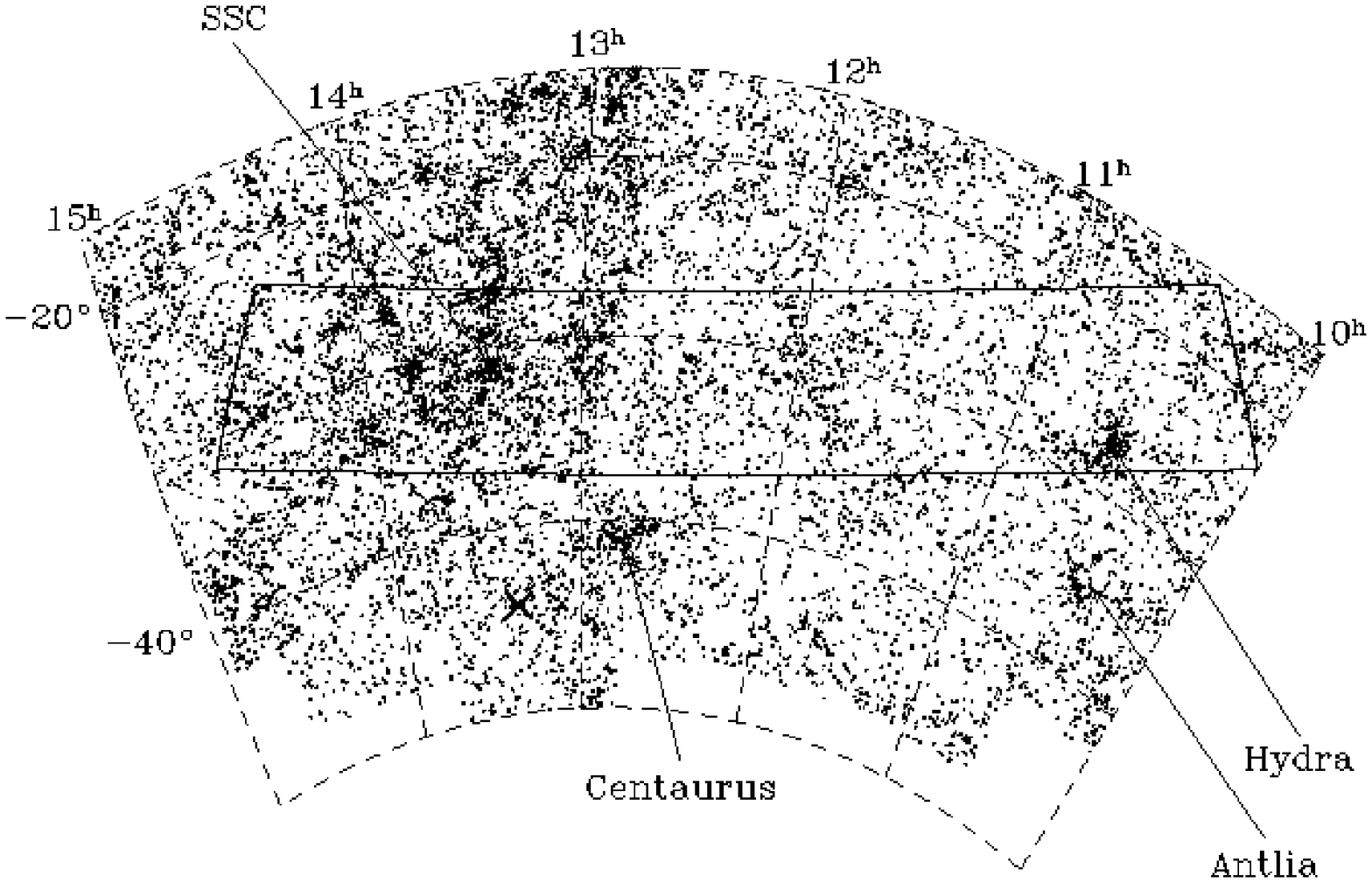,width=0.80\linewidth,angle=-90}
\vspace{-1cm}
\caption{An equal-area Aitoff projection of part of the \hcc\ of
galaxies brighter than $b_J\!=\! 17$ mag, with a
superposed R.A.\ and Dec.\ coordinate grid. The solid line marks the
boundaries of the FLASH survey region, Galactic longitude
$l$=260\degr--330\degr\ and Galactic latitude
$b$=25\degr--35\degr. The \shap\ (SSC) and the Hydra, Centaurus and
Antlia clusters are indicated, and the position of the core of the \GA\
(GA) is marked with an `X'.  The \sgp\  is the obvious over-density
running vertically down the left half of the map. The progressive reduction in galaxy
density towards lower declinations reflects increasing Galactic extinction
towards the Galactic plane.
}
\label{fig:hcc}
\end{figure*}

In this paper we describe the observations and present the survey
catalogue. The arrangement of the paper is as follows:
\S\ref{sec:sample} outlines the selection of the survey sample;
\S\ref{sec:observations} provides the main features of the FLAIR
spectrograph and the observing procedures, \S\ref{sec:reduction} details
the spectroscopic data reduction, \S\ref{sec:redshifts} describes the
measurement of the redshifts and their internal and external errors,
and \S\ref{sec:completeness} gives the completeness and bias of the redshift
sample as a function of field and magnitude. The survey catalogue of
positions, morphological types and redshifts is presented in
\S\ref{sec:catalogue}, while some of the more immediately interesting aspects
of the survey region revealed by the redshift distribution of the
galaxies are discussed in \S\ref{sec:discussion}, and conclusions are
presented in \S\ref{sec:conclusion}. Future papers in this series will
examine the luminosity and correlation functions of the galaxies, and
their variations with morphological type, and investigate the peculiar
velocity field in this volume of space and the effect of the observed
structures on the peculiar motion of the Local Group.

\section{Survey Sample}
\label{sec:sample}

The target catalogue for the FLASH survey, containing \nflair galaxies
brighter than $\bJ$=16.7 (corrected for Galactic extinction), was
drawn from the \hcc\ of
\somak~\shortcite{somak:1990,somak:2002}. The \hcc\ is a
photometric and morphological catalogue of $\sim$20000 galaxies down
to $\bJ$=17 in an extended region of sky in the general direction of
the \GA. Part of the \hcc\ is plotted in an equal-area Aitoff projection in
Figure~\ref{fig:hcc}, which also shows the rectangular $60\degr \times
10\degr$ region of the FLASH survey, bounded by Galactic latitude
$l$=260\degr--330\degr\ and Galactic longitude $b$=25\degr--35\degr.
The FLASH survey region is located just south of the NGP regions of
the Sloan \cite{york:2000} and 2dF \cite{colless:2001} redshift
surveys: it does not overlap with either of them.

The \hcc\ was compiled by scanning 110 UKST Southern Sky Survey plates
(IIIa-J emulsion), using the Automated Photographic Measuring (APM)
facility in Cambridge.  For each object detected on the plate,
the APM measures the position, intensity, and various shape
parameters.  The $(x,y)$ position of each object is converted to RA
and Dec using a six-parameter transformation, with a radial correction
to compensate for the projection effects of the sky onto the Schmidt
plate.  Absolute positions determined in this way are good to
typically about $1\arcsec$.

The APM-measured magnitudes are converted to $\bJ$ magnitudes by
calibration with CCD photometry. Since Schmidt plates overlap
slightly, magnitude calibration from field to field is possible using
the galaxies in the overlap region which, for the \hcc, averages to
approximately 16 galaxies in each overlap region. From these repeat
measurements, the magnitude system on each field can be matched on 
to a common system to better than
$\sim0^m.1$.  It should be noted, however, that bright objects on the
plates may be saturated at the centre, in which case the APM is unable
to produce an accurate magnitude. Objects with magnitudes brighter
than $\bJ = 14.5$ cannot therefore be used in magnitude-sensitive
analyses.

All methods that attempt to distinguish between stars and galaxies
using shape parameters rely on the fact that for stars the images are
more centrally peaked, and described by the point-spread function,
while for galaxies the images are characterised by broader
profiles~\cite{maddox:90}. This method breaks down when the images are
very faint ($\bJ\!>\!20$ on UKST J--plates, since stars and galaxies tend
to look similar), or very bright ($\bJ\!<\! 15$, since the star becomes more
saturated at the centre, while the wings of the stellar profile begin
to approach galaxy-like proportions). The latter effect prevents 
completely automated star-galaxy separation (as in the APM survey,
Maddox et al. 1990) for our sample.

A preliminary star-galaxy separation using the usual shape parameters
like the second-order moments of each image and their areal profiles
\cite{loveday:1996,somak:2002} was performed on each plate with conservative
separation parameters.  The resulting list of galaxy candidates were
then inspected by eye.  A finding chart with ellipses that correspond to
the shape parameters found by the APM simplifies this process, while
also allowing one to inspect how well the APM is detecting all of the
image (important for accurate magnitude determination). During the
inspection process, galaxies were assigned broad morphological
classifications as well (see Table~\ref{tab:catalogue}). 
Further details can be found in \somak~\shortcite{somak:2002}.

In pre--1988 APM scans, due to the nature of the background subtraction
algorithm and the memory allocation during image identification, most galaxies 
larger than $1\arcmin$ were ignored. We have assumed that all
such large galaxies would be in the ESO, PGC and RC3 catalogues, from where
we added all eligible galaxies found in the survey region.
Galaxies added in this way are marked (see Table~\ref{tab:catalogue})--their
positions, diameters and magnitudes 
would have larger uncertainties than the rest of the sample. All galaxies thus
added were visually checked to prevent duplication due to mis-identification
or positional errors.

About 13.5\% of the galaxies in the sample belong to ``merged'' images, where
the image parameters output by the APM refer to a combined image of the
galaxy in question and other overlapping stars or galaxies (classification
label $\ge$81 in Table~\ref{tab:catalogue}. The diameters and magnitudes of these
galaxies will of course be uncertain. We have visually assessed each of these
objects to ensure that the galaxy in question is bright enough to be included
in the sample.

The UK Schmidt plates have generous overlaps with neighbouring plates, since
their centres are 5 degrees apart whereas the usable areas
of these plates are 6.2 degrees square. A relative magnitude system 
is thus established by matching the magnitudes of galaxies in these
overlap regions. CCD observations of galaxy sequences on a large fraction of
of these plates are used to relate it to the $b_J$ magnitude system
\cite{somak:2002}. Since total magnitudes of the calibrating galaxies are used,
the quoted magnitudes refer to the total magnitude of the sample galaxies.

The galaxy magnitudes were corrected for Galactic
extinction according to Burstein \&\
Heiles~\shortcite{burstein:82} reddening values based of H~I column densities,
using a gas-to-dust ratio of $A_B = 4.0*E(B-V)$. The magnitude limit 
of $b_J=16.7$ was imposed {\em after} correction for extinction.

\section{Observations}
\label{sec:observations}

Most of the observations were performed with the FLAIR-II fibre
spectrograph (hereafter FLAIR) system at the UK Schmidt Telescope (UKST)
at Siding Spring Observatory (SSO) in Australia, over a 5-year period
from mid 1991 to early 1996.

The FLAIR system \cite{pw:1994} was chosen as the primary observing
instrument because it best matched the requirements of the FLASH survey.
With $\sim$90 fibres across a Schmidt field (40~deg$^2$), the FLAIR
system is well-suited to large area, relatively bright surveys
($B\leq18$), and is most efficient when the number density of objects is
in the range 1--10~deg$^{-2}$. The FLASH survey contains \nflair objects
across 605~deg$^2$ of sky, and thus is ideally suited to an instrument
like FLAIR. Although other fibre-spectrograph systems on larger
telescopes have more fibres, they also have much smaller fields of view,
and so are inefficient compared to FLAIR when the surface density of
objects is low and the survey region covers a large area of sky.

With the FLAIR system, it is natural to tile the survey region using the
the UKST Sky Survey fields, since the fibres are directly positioned
onto the images of the target galaxies on Sky Survey copy plates. For
each Schmidt field, a target list containing 90--100 galaxies was
generated from the survey catalogue. If the field is densely populated,
further target lists were created and the field re-observed. Once
selected, the target galaxies were marked on UKST Sky Survey plate
copies, along with at least 5 fiducial stars scattered evenly around the
field, and a suitable guide star near the centre of the field.

Although the observer is assisted by a robotic system known as
AutoFred~\cite{bedding:1993}, the fibering process is essentially
performed by hand. A small monitor allows the volunteer to see a
close-up of the Schmidt plate around each target galaxy, and to mark the
centre of the galaxy's image. A `button' holding a right-angle prism
feeding a 100$\mu m$ (6\farcs7) diameter fibre is then attached to the
robotic gripper arm. A carefully measured amount of bonding agent is
placed on the base of the button, and the robotic arm is lowered to the
plate. The fibre bundle is back-illuminated by a small lamp, which
allows a computer to estimate the centroid of the fibre. Minute manual
adjustments are made to the position of the gripper until the fibre
centroid matches the position of the galaxy marked earlier. This process
is quite quick, and results in the fibre being positioned onto the
target galaxy to within the required $10\mu m$ (0\farcs7) accuracy.
The fibre is then illuminated with UV light for 25s, which cements the
fibre button onto the plate.

As many fibres as possible are placed onto target galaxies, with the
intention of achieving a sampling rate close to 1-in-1. In those cases
where significant over-densities existed in the field, a second or third
fibre configuration for the field was observed.

As with any multi-fibre system, the finite size of the button limits how
close together one can place the fibres on the plate. This can be a
problem in the over-dense regions, although multiple observations of
densely-populated fields, coupled with the flexibility of the FLAIR
system with respect to positioning the fibres, reduce the severity of
this problem. The size of the gripper limits how close to the field edge
one can place a fibre, and a small $11\arcmin \times 199\arcmin$
rectangular area from the centre of the field to the bottom of the plate
is obscured by the guide-star fibre bundle.

Around 6--10 fibres are reserved for observing sky. These are placed
evenly across the plate on image-free patches of sky, although for
improved sky subtraction it is also a good idea to place more sky fibres
near densely-placed fibres. A small coherent fibre bundle, placed on a
bright star close to the centre of the field, is used for coarse
acquisition of the field and for guiding. Another 5 fibres, with thinner
55$\mu m$ cores, are placed on moderately bright stars scattered around
the field, used as astrometric fiducials in the fine acquisition
and alignment of the field.

Once fibering is complete, the plate is tensioned to obtain the correct
curvature for the Schmidt telescope focal plane, the plateholder is
loaded into the telescope, and the fibre bundle is connected to the
spectrograph. The guide star is used to find the centre of the field, and
the fiducial stars are used to acquire the correct rotation of the field.
They are also used to help track the field across the sky during the night.

The survey observations used the 300B grating (300~lines/mm, blazed in
the blue), giving an observable wavelength range from about 4000\AA\ to
7000\AA, and a resolution of about 5\AA/pixel (FWHM$\approx$2.5\AA). This
setup allowed the detection of the 4000\AA\ break down to almost zero
redshift, and the detection of H$\alpha$ in emission out to redshifts of
$z \sim 0.1$.

Individual exposures were typically 2000$s$ long, and up to 10 exposures
of a single field could be taken in a clear
half-night. (The FLAIR system could only observe one or two fields per
night due to the limited number of fibre plate-holders and the long time
required for manual fibre configuration.) Individual exposures were
combined during the reduction process to remove cosmic rays and to
improve the signal-to-noise ratio of the spectra.

\section{Data Reduction}
\label{sec:reduction}

Data reduction for the spectroscopy consists of extracting each
individual one-dimensional fibre spectrum from the two-dimensional data
frame, calibrating the wavelength scale of the spectra, and subtracting
the sky background. All these activities were performed with the {\sc
flair} data reduction package in {\sc iraf} \cite{holman:94}.

Bias frames were taken each night. Where possible, sky flats were taken
for fibre-throughput calculations; otherwise dome flats were used. Arc
exposures for wavelength calibration were taken before and after the
observations of each field. Two types of arcs were used to produce
sufficient numbers (20--25) and coverage of emission lines across the
wavelength range of interest: Ne for the red end and Hg-Cd for the blue.
No discernible wavelength shifts were ever measured between the arc
exposures taken half a night apart on either side of the survey
exposures. This is due to the fact that the spectrograph in the FLAIR
system is mounted on an optical bench on the floor of the dome rather
than on the telescope itself, thereby avoiding flexing of the
spectrograph unit as the telescope follows the target field across the
sky.

The object exposures were combined to form a single object image, which
was then processed by the {\sc iraf} task {\sc dohydra}. This task uses
the combined sky-flat image to identify the position of the spectra in
the object image, although it does require manual intervention at times,
as the spectra on the object image are 1--3 pixels wide with only a
$\sim2$ pixel separation. The relatively even illumination across the
twilight sky image ensures that the spectra are correctly traced from
the red wavelengths to the blue. Third-order polynomials are quite
adequate to fit the trace of the spectra, with typical rms errors being
0.1--0.2 pixels.

The combined flat-field frames were then used to calculate the relative
fibre throughput, for which compensations are made in the extracted
spectra. Fibre throughput tends to degrade with time as dust and grime
accumulate on the prisms. As the system ages, the glue between the
prism and fibre becomes more opaque and accidental kinks in the fibres
take their toll. Misalignment in the mounting of the prism on the fibre
buttons due to maintenance repairs or knocks in positioning, and poor
seating of the fibre button the plate during configuration, also reduce a
fibre's throughput, as do telescope vignetting and the variation in
transmission of a fibre from the manufacturing process.

The wavelength calibration is based on identifying 20--25 lines in the
arc spectra and fitting these with third- or fourth-order polynomials to
obtain the wavelength--pixel relation. Typical rms errors in the fits
are about 0.1~pixels, or about 0.5\AA.

The individual sky spectra are then inspected, and any spectrum
contaminated by light bleeding from bright objects in neighbouring
fibres is removed before the sky spectra are combined. Although
potentially a serious problem, contamination from neighbouring fibres is
quite rare, and tends to manifest as a modification of the spectral
continuum of both objects and sky. This does not affect the position of
spectral lines, but only the strength of the lines relative to the
continuum; for measuring redshifts it is therefore not a particular
problem except where the sky spectrum, and hence sky subtraction, is
affected. There is often considerable variation in the sky spectra due
to the large field of view of the Schmidt telescope, which results in
the sky fibres sampling parts of the sky up to 6\degr\ apart. Most of
the variations are in the strengths of the strong \hbox{O\,{\sc i}}
emission features at 5577\AA, 6300\AA\ and 6363\AA, while the continuum
tends to remain unaffected. This results in good continuum subtraction,
but poor subtraction of the strongly-varying sky lines, which can lead
to difficulties in identifying spectral features near these lines.

\section{Redshifts}
\label{sec:redshifts}

\subsection{Redshift Measurements}
\label{ssec:measurements}

Once the spectra have been extracted, sky-subtracted, and
wavelength-calibrated, radial velocity (redshift) measurements can be
made. The CCD on the FLAIR system prior to 1995 had a rather poor
response in the blue, making it difficult to identify the Ca H and K
lines and the G band. While the new thinned CCD is twice as responsive
in the blue, these lines can still be quite difficult to detect. Other
lines used in the process of measuring the redshift were H$\beta$ at
4861\AA, \hbox{[O\,{\sc iii}]} at 4959\AA\ and 5007\AA, the Mg~b triplet
around 5175\AA, the Fe features at 5207\AA\ and 5270\AA, Na~D at
5892\AA, H$\alpha$ at 6563\AA, \hbox{N\,{\sc ii}} at 6584\AA, and the
\hbox{S\,{\sc ii}} doublet at 6716\AA\ and 6731\AA.

Using the {\sc iraf} task {\sc rvidlines}, gaussians were fitted to each
identified feature in the spectrum to determine the position of the line
and hence its shift. The line shifts are then averaged to produce a
final radial velocity measurement, while the standard deviation of the
line shifts gives the internal radial velocity error. These errors were
typically less than 100~\kms\ (see below). 

Additional redshifts for the FLASH survey catalogue were obtained from
the literature via the NASA/IPAC Extragalactic Database (NED) and the
ZCAT catalogue~\cite{zcat}; 901 redshifts were obtained from ZCAT, and
354 were obtained from NED.

The final estimate of the galaxy's redshift is the variance-weighted
mean of the measured redshifts from FLAIR (including repeat measurements
for some objects) and the literature. A further, subjective, weighting
was applied to the literature-derived redshifts, reflecting our degree
of belief in the estimated redshift errors in the literature---in short,
NED and ZCAT redshifts were given half the weight of the FLAIR
redshifts.

\subsection{Internal and External Errors}
\label{ssec:zerrors}

\begin{figure}
\epsfig{file=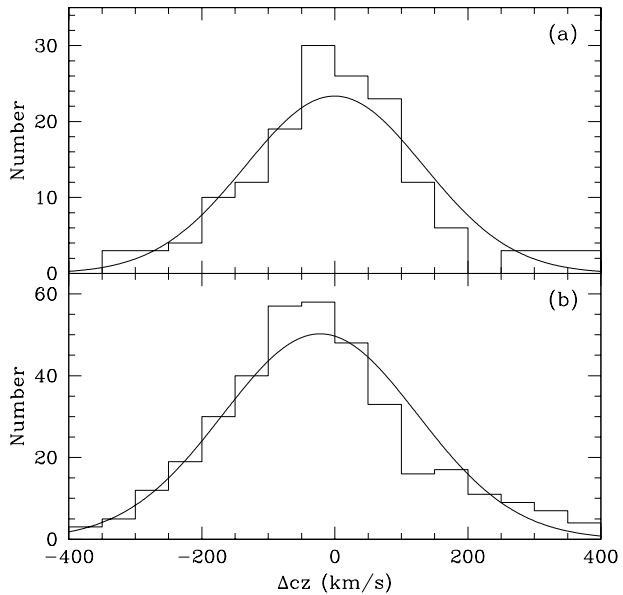,width=\linewidth}
\caption{Comparing redshifts measured on the FLAIR system with (top)
repeat FLAIR measurements, and (bottom) literature redshifts. Solid
curves are gaussians with the median and standard deviation about the
median of each distribution: $-$0.4~\kms\ and 134~\kms\ for the top panel,
and $-$22~\kms\ and 146~\kms\ for the bottom panel.}
\label{fig:deltav}
\end{figure}

One can check the precision of the FLAIR redshifts by comparing the
multiple radial velocity measurements for those galaxies that have
repeat observations. The redshift differences for the 164 galaxies with
multiple redshift measurements are plotted in the top panel of
Figure~\ref{fig:deltav}. The mean offset is $-$0.4~\kms; the standard
deviation in the difference is 134~\kms, indicating an rms error in the
redshifts of 95~\kms\ (assuming equipartition of errors). 

Although we have been careful in estimating the redshift errors, there
may be effects (systematic or otherwise) that have not been accounted
for. To detect such effects and determine the reliability of the
estimated errors, one can compare the variance-weighted estimated
errors, $\overline{\sigma}$, with the rms errors, $\sigma_{\rm RMS}$, for
the 164 objects with repeat measurements.

\begin{figure}
\epsfig{file=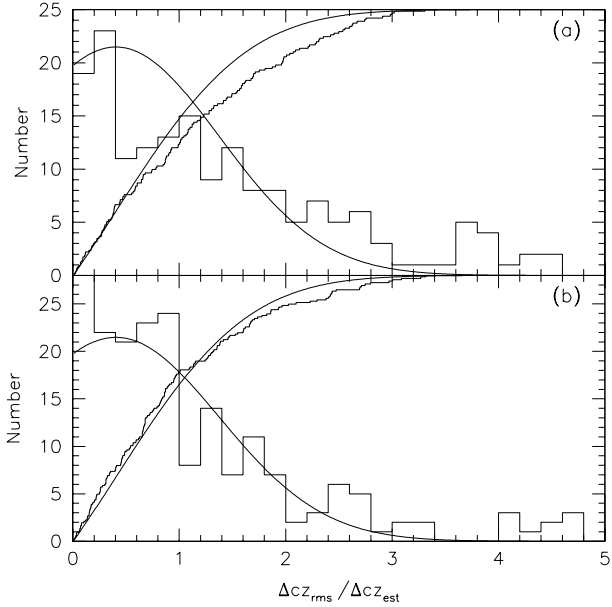,width=\linewidth}
\caption{Comparison of estimated errors to rms errors from repeat
measurements. The top panel shows the comparison using the raw estimated
errors; the bottom panel shows the comparison using the corrected
estimated errors.}
\label{fig:errnorm}
\end{figure}

The top panel of Figure~\ref{fig:errnorm} shows both the differential
(histogram) and cumulative (stepped line) distributions of the ratio
$\sigma_{\rm RMS}/\overline{\sigma}$ for all the objects with multiple FLAIR
redshift measurements. The corresponding smooth curves are the predicted
differential and cumulative distribution functions for this quantity
assuming the estimated errors are correct and the error distributions
are gaussian \cite{colless:1999}. The top panel
of Figure~\ref{fig:errnorm} shows that the errors are under-estimated,
since, from the cumulative distributions, it is evident that the ratio
of the rms to estimated errors is always greater than expected.

This under-estimation of the errors in the FLAIR redshifts results from
the way the errors were estimated from the spectra. The {\sc rvidlines}
algorithm finds the rms error of the gaussian fits to the individual
line profiles, and so produces an {\em internal} error; however,
external errors (such as wavelength calibration errors, mis-alignment of
the fibre and the centre of the galaxy, or miscellaneous instrumental
offsets) can also contribute to the overall error in the measured
redshift.

We find empirically that adding 40~\kms\ in quadrature to the estimated
errors optimises the match between the estimated errors and the rms
errors (maximising the probability that the predicted and observed
distributions of $\sigma_{\rm RMS}/\overline{\sigma}$ are the same in a \ks\
test). The resulting distribution of $\sigma_{\rm RMS}/\overline{\sigma}$ is
shown in the lower panel of Figure~\ref{fig:errnorm}, where the improved
agreement is quite evident. The median rms error (re-calculated allowing
for this correction) is 103~\kms. This is consistent with the results of
Watson \shortcite{watson:91}, who showed that the FLAIR system could
reproduce redshifts to better than 150~\kms\ for $b_J\!\sim\! 17$ galaxies.

We can also obtain an external estimate of the errors in the FLAIR
redshifts using the 393 objects with both FLAIR and literature
redshifts. The lower panel of Figure~\ref{fig:deltav} shows the
distribution of the difference in redshifts between FLAIR and literature
measurements, after re-computing the variance-weighted FLAIR redshifts
using the corrected estimated errors. The mean offset is $-$22~\kms\ with
a standard deviation of 146~\kms, which, with equipartition of errors,
implies an rms redshift error of 103~\kms, in excellent agreement with the
corrected estimated errors.

\section{Redshift Completeness and Bias}
\label{sec:completeness}

\begin{table}
\caption{Statistical properties of the Schmidt fields in the FLASH survey region.}
\label{tab:fields}
\begin{tabular}{cccccl}
(1) &  (2) & (3) & (4) & (5) & (6) \\
Field & \multicolumn{2}{c}{Field Center} & $\rm N_{gal}$ & $\rm N_{cz}$ & Completeness \\
 & RA & Dec & & \\
322 & 12 34  & --40 00&    1 &    1 &  1.00  \\
323 & 13 00  & --40 00&    3 &    1 &  0.33  \\
377 & 11 12  & --35 00&   26 &    4 &  0.15  \\
378 & 11 36  & --35 00&   90 &   50 &  0.56  \\
379 & 12 00  & --35 00&   90 &   61 &  0.68  \\
380 & 12 24  & --35 00&  118 &   90 &  0.76  \\
381 & 12 48  & --35 00&  150 &   97 &  0.65  \\
382 & 13 12  & --35 00&  220 &  173 &  0.79  \\
383 & 13 36  & --35 00&  253 &  215 &  0.85  \\
384 & 14 00  & --35 00&  175 &  133 &  0.76  \\
385 & 14 24  & --35 00&   40 &    9 &  0.22  \\
436 & 10 21  & --30 00&    9 &    9 &  1.00  \\
437 & 10 44  & --30 00&  133 &   68 &  0.51  \\
438 & 11 07  & --30 00&   97 &   62 &  0.64  \\
439 & 11 30  & --30 00&   79 &   66 &  0.84  \\
440 & 11 53  & --30 00&  243 &  127 &  0.52  \\
441 & 12 16  & --30 00&  107 &   76 &  0.71  \\
442 & 12 39  & --30 00&   94 &   70 &  0.74  \\
443 & 13 02  & --30 00&  341 &  280 &  0.82  \\
444 & 13 25  & --30 00&  437 &  347 &  0.79  \\
445 & 13 48  & --30 00&  328 &  234 &  0.71  \\
446 & 14 11  & --30 00&  203 &  120 &  0.59  \\
447 & 14 34  & --30 00&  225 &  109 &  0.48  \\
499 & 09 54  & --25 00&    2 &    1 &  0.50  \\
500 & 10 16  & --25 00&   75 &   59 &  0.79  \\
501 & 10 38  & --25 00&  185 &  158 &  0.85  \\
502 & 11 00  & --25 00&   85 &   47 &  0.55  \\
503 & 11 22  & --25 00&   61 &   37 &  0.61  \\
504 & 11 44  & --25 00&   49 &   30 &  0.61  \\
505 & 12 06  & --25 00&   28 &    5 &  0.18  \\
506 & 12 28  & --25 00&    1 &    1 &  1.00  \\
508 & 13 12  & --25 00&   14 &   11 &  0.79  \\
509 & 13 34  & --25 00&   85 &   42 &  0.49  \\
510 & 13 56  & --25 00&  172 &  109 &  0.63  \\
511 & 14 18  & --25 00&  141 &   83 &  0.59  \\
512 & 14 40  & --25 00&   15 &    5 &  0.33  \\
567 & 10 09  & --20 00&   49 &   36 &  0.73  \\
568 & 10 30  & --20 00&  106 &   84 &  0.79  \\
569 & 10 51  & --20 00&   52 &   27 &  0.52  \\
570 & 11 12  & --20 00&    5 &    0 &  0.00  \\
638 & 10 20  & --15 00&   19 &    2 &  0.11  \\
639 & 10 40  & --15 00&    7 &    2 &  0.29  \\
\end{tabular}

\medskip
(1) Schmidt field ID; (2) Field centre right ascension (B1950); 
(3) Field centre declination (B1950); (4) Number of galaxies in the
field belonging to our sample; 
(5) Number of galaxies with redshifts; (6) Field completeness
(percentage of redshifts measured).
\end{table}

Incompleteness within a survey can occur in several ways, and for many
reasons. If an incompleteness is {\em systematic} with respect to an
observable quantity, it can seriously affect any analysis performed on
the survey data if it is uncorrected. Incompleteness can be corrected
for when it relates to a quantity that is known for every object in the
input catalogue (\eg\ magnitude or morphological type), but not if it is
due to a quantity that is being measured (\eg\ redshift). To correct for
an incompleteness, one may simply assign an appropriate weight to each
galaxy, although for some analyses, a simple weighting strategy will not
work, and the incompleteness must be factored into the analysis directly.

The overall redshift completeness of the FLASH survey is 0.64, with
redshifts for \nz of the \nflair galaxies in the catalogue. We now
consider the dependence of this completeness on observing conditions
(field), on apparent magnitude, and on redshift itself.

\subsection{Field Completeness}
\label{ssec:field_comp}

\begin{figure*}
\epsfig{file=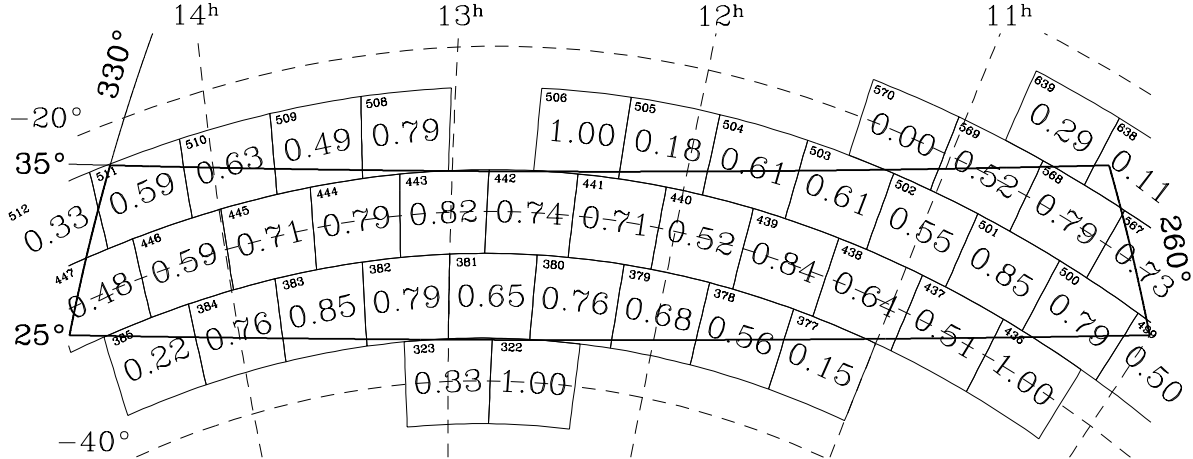,width=0.40\linewidth,angle=-90}
\caption{Variation in redshift completeness from field to field in the FLASH
survey.}
\label{fig:fieldcomp}
\end{figure*}

Because we observe galaxies with FLAIR one Schmidt plate at a time, there
is a field-to-field variation in completeness that depends on the quality
of the observing conditions for each field.  Figure~\ref{fig:fieldcomp}
shows the Schmidt fields that tile the FLASH survey region, with the field
number shown in the top left and the field redshift completeness centred on
each field.

The redshift completeness across the survey region is reasonably uniform
given the variation in galaxy number density across the sky in the region
(see Figure~\ref{fig:hcc}). The median completeness over all 42 fields is
0.63. The reason this figure is lower than the overall completeness is that
the lower-completeness fields tend to be near the edge of the survey
region, where the number of galaxies within the survey is too small to be
able to use the FLAIR system efficiently, so that galaxy redshifts in these
fields are predominantly from the literature.

\begin{figure}
\epsfig{file=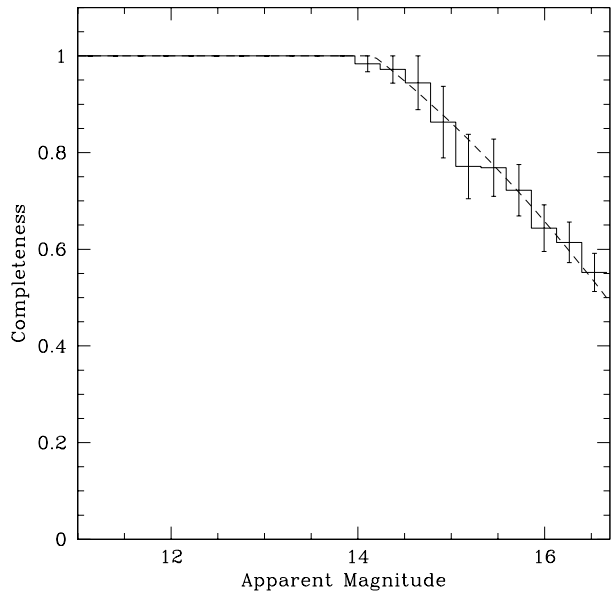,width=\linewidth}
\caption{Redshift completeness as a function of (corrected) apparent magnitude for all
galaxies in the FLASH survey. The dashed line is the fit used in
applying the magnitude-dependent completeness correction of the data:
$f(m) = -0.96 + 0.43m - 0.020m^2$ for apparent magnitudes $m > 14.2$, and
$f(m) = 1$ otherwise.}
\label{fig:mag_comp_all}
\end{figure}

\begin{figure}
\epsfig{file=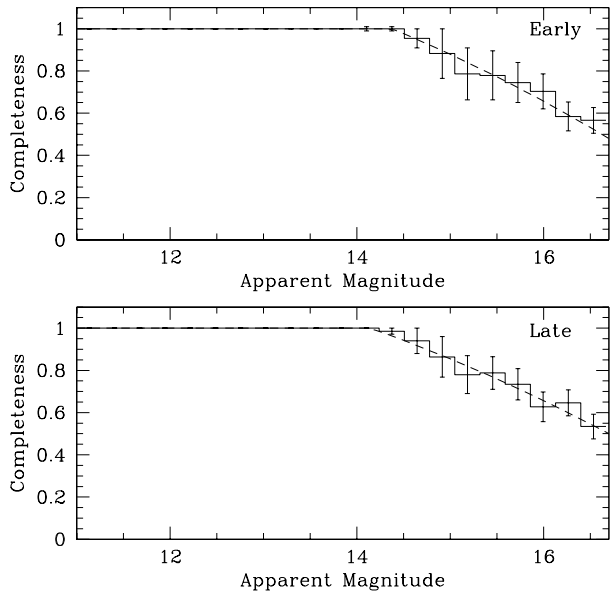,width=\linewidth}
\caption{Redshift completeness as a function of (corrected) apparent magnitude for early and
late morphological types in the FLASH survey. The dashed line is the fit
used in applying the magnitude-dependent completeness correction of the
data. For early types: $f(m) = 0.00088 + 0.32m - 0.018m^2$ for $m > 14.2$,
unity otherwise. For late types: $f(m) = 0.21 + 0.27m - 0.015m^2$, for
$m > 14.2$, unity otherwise.}

\label{fig:mag_comp_sub}
\end{figure}

Table~\ref{tab:fields} lists the statistics of each Schmidt field in the
survey region. For each field the table contains the following
information: field ID number, R.A. and Dec.\ of the field centre
(B1950), number of target galaxies, number of redshifts obtained, and
field completeness.

\subsection{Apparent Magnitude Completeness}
\label{ssec:app_mag_comp}

The signal-to-noise ratio in the observed spectra decreases as the
target objects' apparent magnitudes become fainter. This results in a
decreasing redshift completeness at fainter apparent magnitudes which,
if not corrected for, will result in an incorrect selection function and
skew the results obtained from magnitude-dependent analyses (\eg\ the
luminosity function).

The completeness as a function of apparent magnitude for the whole
survey sample is shown in Figure~\ref{fig:mag_comp_all}, and for early
and late morphological types separately in
Figure~\ref{fig:mag_comp_sub}. Both figures show a rapid drop in
completeness beyond $\bJ$=15, with approximately 50 per cent
completeness at the survey's magnitude limit. Also shown in the figures
are the best-fit polynomial functions approximating the completeness
variation over the magnitude range (dashed lines). These functions are
used in applying the magnitude-dependent completeness corrections in
later papers.

\subsection{Redshift Bias}
\label{ssec:bias}

In early May and mid April 1994, 34 redshifts were obtained with the
Double Beam Spectrograph (DBS) on the ANU 2.3m telescope at SSO. The
target galaxies were a selection of objects observed with FLAIR for
which redshifts were {\em not} obtained. The purpose of these further
observations was to explore the possibility that the FLASH survey might
be biased against obtaining redshifts for some subset of objects in the
target catalogue on the basis of their spectral type or redshift.

\begin{figure}
\epsfig{file=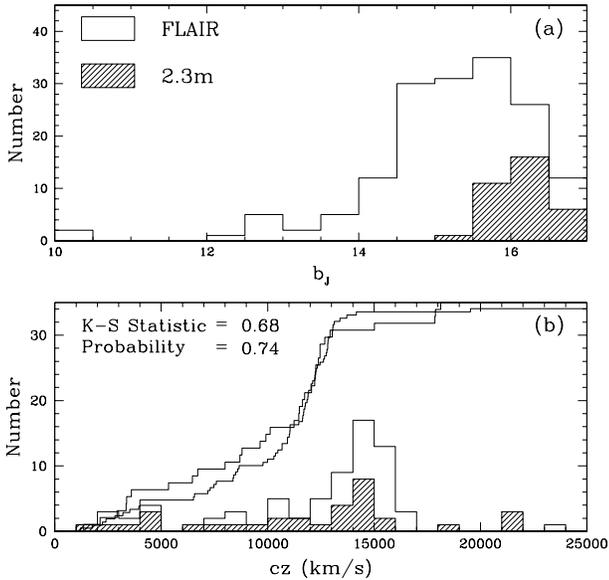,width=\linewidth}
\caption{Comparison of redshift distributions for objects with and
without FLAIR redshift measurements. Redshifts for the latter were
obtained with the ANU 2.3m telescope. The top panel shows the magnitude
distributions of the two samples. The bottom panel shows the
differential and cumulative redshift distributions.}
\label{fig:ks}
\end{figure}

The top panel of Figure~\ref{fig:ks} shows the number--magnitude
distribution of galaxies for which redshifts were obtained with FLAIR or
the ANU 2.3m in the same fields. Clearly the objects that failed to have
a redshift measured with FLAIR (the ANU 2.3m sample) are at the faint
end, with typical apparent magnitudes fainter than $\mbJ$=15.5,
consistent with Figure~\ref{fig:mag_comp_all}.

The bottom panel of Figure~\ref{fig:ks} shows the number--redshift
distribution of the FLAIR and 2.3m samples, along with the cumulative
distributions from each sample. The number--redshift distributions for
the FLAIR sample and 2.3m sample appear quite consistent. We use the
\ks\ statistic to quantitatively test the hypothesis that the galaxies
with and without FLAIR redshifts come from the same underlying redshift
population. We find that the two redshift distributions are consistent
at the 74 per cent confidence level.

We can conclude, therefore, that the galaxies observed with FLAIR for
which no redshifts were obtained have the same redshift distribution as
those for which we obtained FLAIR redshifts. The only redshift bias in
the FLASH survey is that due to magnitude-dependent incompleteness, and
can therefore be corrected in a straightforward manner.

\section{The FLASH Survey Catalogue}
\label{sec:catalogue}

The FLASH survey includes a total of \nz redshifts over 605~deg$^2$ of
sky. An example subset of the FLASH survey catalogue is presented in
Table~\ref{tab:catalogue}.
The full survey catalogue is available from NASA's
Astrophysical Data Centre (ADC; http://adc.gsfc.nasa.gov/adc.html)
and the Centre de Donn\'{e}es astronomiques de Strasbourg
(CDS; http://cdsweb.u-strasbg.fr/).
Each entry in the table contains: the Schmidt field the
galaxy was observed in, the galaxy's R.A.\ and Dec.\ (B1950), the
magnitude ($\bJ$), major axis diameter (arcmin), morphological type,
heliocentric radial velocity and error (\kms), and source reference for
the radial velocity (FLAIR, 2.3m, NED, ZCAT). The magnitudes were
extinction-corrected according to
Burstein \&\ Heiles~\shortcite{burstein:82}, using $A_B =
4.0*E(B-V)$.

\begin{figure*}
\epsfig{file=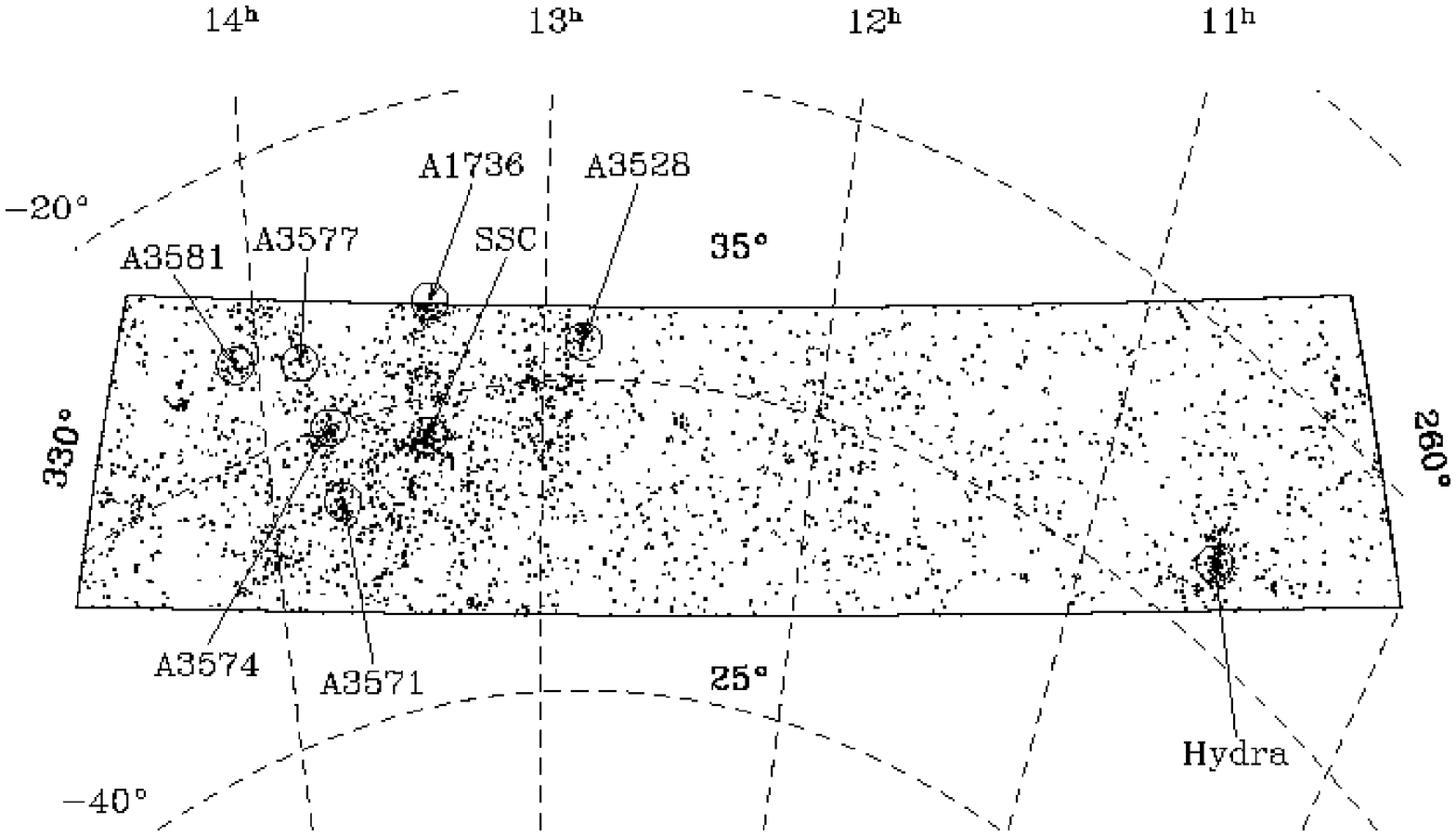,height=15.0cm, angle=-90}
\caption{The distribution of the FLASH survey galaxies with measured
redshifts on the sky, shown
in an equal-area Aitoff projection. 
Some of the more
prominent clusters within the survey region are indicated. The ``SSC'' refers
to the four clusters in the core of the Shapley Supercluster, 
consisting of the A3558, SC1327-312, SC1329-313 and A3562 clusters.}
\label{fig:skyzplot}
\end{figure*}

Figure~\ref{fig:skyzplot} displays those galaxies in the FLASH survey
catalogue with measured redshifts in an equal-area Aitoff projection.
Clusters and filamentary structures are clearly evident across the survey
region: the \shap\ dominates the survey region near
$(l,b)$=(312\degr, 31\degr); the Hydra cluster of galaxies is clearly
visible at the lower right of the plot; a number of other prominent Abell
clusters are indicated on the figure. The flattened structure spanning the
survey region north-south at about $13^h$ (see also Figure~\ref{fig:hcc})
is the \sgp, the projection onto the sky of the Local Supercluster.

\begin{table*}
\begin{minipage}{120mm}
\caption{The FLASH survey catalogue -- example data.
The full survey catalogue will be available from NASA's
Astrophysical Data Centre (ADC; http://adc.gsfc.nasa.gov/adc.html)
and the Centre de Donn\'{e}es astronomiques de Strasbourg
(CDS; http://cdsweb.u-strasbg.fr/).}
\label{tab:catalogue}
\begin{tabular}{lccccrrrrrc}
(1) &  (2) & (3) & (4) & (5) & (6) & (7) & (8) & (9) & (10) & (11) \\
Field &  RA & Dec & $\ell$ & b & $\bJ$ & $D$ & $T$ & $cz$ & $\sigma$ & cz \\
 &  \multicolumn{2}{c}{(B1950)} & (\degr) & (\degr) & (mag) & (\arcmin) & & \multicolumn{2}{c}{(\kms)} & src \\

499 & 10 03 24.84 & -23 38 21.8 & 261.0 & 25.2 & 16.6 & 0.5 &  1 &     0 &   0 &       - \\
499P& 10 03 36.00 & -22 48 48.0 & 260.4 & 25.8 & 14.5 & 1.1 &  2 &  3829 &  39 &       F \\
567 & 10 07 22.57 & -22 02 43.3 & 260.6 & 27.0 & 16.7 & 0.6 &  3 & 14005 &  58 &       F \\
500P& 10 07 43.00 & -24 04 59.9 & 262.1 & 25.5 & 14.6 & 1.2 &  3 &  9565 &  27 &       N \\
567 & 10 07 59.20 & -21 10 31.3 & 260.1 & 27.7 & 16.7 & 0.3 &  4 & 16740 &  45 &       F \\
500 & 10 08 13.14 & -23 43 56.7 & 262.0 & 25.8 & 15.8 & 0.5 &  4 &  9699 & 112 &       F \\
500 & 10 08 31.25 & -22 54 40.5 & 261.5 & 26.5 & 15.8 & 1.0 &  3 &  9689 & 190 &       F \\
567 & 10 08 40.86 & -22 26 04.8 & 261.1 & 26.9 & 15.3 & 0.8 &  1 &  3665 & 111 &       F \\
567 & 10 08 46.35 & -21 17 05.3 & 260.3 & 27.8 & 15.4 & 0.7 &  3 &  9323 &  10 &     F,Z \\
567 & 10 09 07.76 & -21 03 01.1 & 260.2 & 28.0 & 16.0 & 0.4 &  3 &  9110 &  70 &       F \\
567 & 10 09 14.69 & -20 54 37.0 & 260.2 & 28.1 & 16.3 & 0.7 & 84 &     0 &   0 &       - \\
567P& 10 09 15.54 & -21 00 23.6 & 260.2 & 28.1 & 14.2 & 0.7 &  3 &  8972 &  10 &     N,Z \\
567 & 10 09 17.71 & -21 07 11.5 & 260.3 & 28.0 & 15.9 & 0.5 &  4 &  9274 &  54 &       F \\
500 & 10 10 43.88 & -23 12 17.9 & 262.1 & 26.6 & 16.5 & 0.5 &  3 & 15593 &  70 &       F \\
500P& 10 11 13.12 & -22 30 35.3 & 261.7 & 27.2 & 15.3 & 0.9 &  3 &  3617 &  32 &       F \\
500 & 10 11 15.32 & -23 15 01.4 & 262.2 & 26.6 & 16.4 & 0.4 &  3 & 23911 &  75 &       F \\
500P& 10 11 40.01 & -25 23 30.0 & 263.8 & 25.0 & 16.0 & 1.2 &  3 &     0 &   0 &       - \\
567P& 10 11 42.00 & -21 43 42.0 & 261.2 & 27.9 & 14.1 & 2.2 &  3 &  3615 &   6 &     F,N \\
567 & 10 12 00.73 & -20 08 42.3 & 260.2 & 29.1 & 15.7 & 0.9 & 83 &  3465 & 109 &       F \\
500 & 10 12 08.97 & -22 48 04.9 & 262.1 & 27.1 & 16.0 & 0.5 &  2 &  5094 &  35 &       F \\
567 & 10 12 18.36 & -20 45 42.7 & 260.7 & 28.7 & 15.9 & 0.8 &  3 &     0 &   0 &       - \\
567P& 10 12 21.19 & -20 33 40.8 & 260.5 & 28.9 & 15.2 & 1.2 &  3 &  3595 &  51 &       F \\
567 & 10 12 30.68 & -20 42 55.5 & 260.7 & 28.8 & 16.3 & 0.8 &  3 &     0 &   0 &       - \\
567 & 10 12 31.26 & -20 30 16.2 & 260.5 & 28.9 & 16.2 & 0.4 &  4 &     0 &   0 &       - \\
500 & 10 12 32.51 & -22 48 00.2 & 262.2 & 27.2 & 15.2 & 1.0 &  1 &  3679 &  25 &     F,N \\
500 & 10 12 45.93 & -24 57 13.5 & 263.7 & 25.5 & 16.6 & 0.5 &  3 & 11740 & 108 &       F \\
567 & 10 13 01.52 & -22 18 41.4 & 261.9 & 27.6 & 15.8 & 0.6 & 83 &     0 &   0 &       - \\
567 & 10 13 10.01 & -20 20 04.4 & 260.5 & 29.2 & 15.7 & 0.6 &  1 &     0 &   0 &       - \\
567 & 10 13 12.13 & -19 45 59.1 & 260.1 & 29.6 & 16.2 & 0.5 & 82 & 14423 &  60 &       F \\
567P& 10 13 14.19 & -20 23 52.8 & 260.6 & 29.1 & 13.6 & 1.8 &  2 &  3565 &  35 &     F,N \\
567 & 10 13 21.67 & -20 02 46.1 & 260.4 & 29.4 & 14.0 & 1.6 &  3 &  3653 &  22 &     F,N \\
567 & 10 13 25.19 & -21 29 12.4 & 261.4 & 28.3 & 15.4 & 0.9 & 83 &     0 &   0 &       F \\
567 & 10 13 30.68 & -21 24 57.5 & 261.4 & 28.4 & 15.2 & 0.7 &  4 &  3547 &  54 &       F \\
567 & 10 13 33.61 & -19 47 28.4 & 260.2 & 29.6 & 16.5 & 0.4 & 73 & 16413 &  42 &       F \\
567 & 10 13 38.65 & -21 26 19.1 & 261.4 & 28.4 & 16.7 & 0.6 &  4 &     0 &   0 &       - \\
567 & 10 13 51.10 & -20 20 50.8 & 260.7 & 29.3 & 16.5 & 0.4 &  4 & 12964 &  75 &       F \\
567 & 10 13 58.29 & -19 56 22.6 & 260.4 & 29.6 & 15.7 & 0.8 &  2 &     0 &   0 &       - \\
567 & 10 14 07.76 & -19 37 27.8 & 260.2 & 29.9 & 15.8 & 0.7 & 73 & 14925 &  90 &       F \\
500 & 10 14 20.22 & -25 14 21.3 & 264.2 & 25.5 & 16.7 & 0.4 &  2 & 11894 &  64 &       F \\
567 & 10 14 27.34 & -20 23 09.8 & 260.8 & 29.3 & 15.3 & 0.8 &  3 &  7046 &  80 &       F \\
500 & 10 14 30.21 & -25 01 06.6 & 264.1 & 25.7 & 15.7 & 0.5 &  3 & 12042 & 110 &       F \\
567 & 10 14 32.05 & -21 58 48.5 & 262.0 & 28.1 & 15.4 & 0.7 & 81 &  3932 &  40 &       F \\
500 & 10 14 37.62 & -24 55 51.0 & 264.1 & 25.8 & 15.8 & 0.6 &  2 & 12627 &  75 &       F \\
567 & 10 14 50.66 & -20 48 59.7 & 261.2 & 29.0 & 15.3 & 0.7 &  2 &  3478 &  31 &       F \\
500 & 10 15 18.20 & -24 16 44.5 & 263.8 & 26.4 & 16.4 & 0.6 &  1 & 12289 & 108 &       F \\
500 & 10 15 20.37 & -26 02 02.0 & 265.0 & 25.0 & 16.3 & 0.4 & 81 &     0 &   0 &       - \\
500 & 10 15 33.43 & -24 18 39.8 & 263.8 & 26.4 & 16.2 & 1.0 &  3 & 12725 & 108 &       F \\
567 & 10 15 56.20 & -22 09 48.6 & 262.4 & 28.2 & 15.3 & 0.9 &  3 &  3975 &  42 &       F \\
\end{tabular}

\medskip
(1)~The
Schmidt field (``P'' indicates that the position was taken from the PGC or RC3
catalogue); (2)~The galaxy's right ascension (B1950); (3)~The
galaxy's declination (B1950); (4)~Galactic longitude; (5)~Galactic
latitude; (6)~$\bJ$ magnitude; (7)~Major axis diameter (arcmin);
(8)~Morphological type flag [1: Elliptical, 
2: S0,
72: SB0,
3 : Spiral,
73: Barred Spiral,
4 : Late-type Irregular,
5 : Unknown,
81: Star + Elliptical,
82: Star + S0,
83: Star + Spiral,
84: Star + Irregular,
91: Elliptical + Elliptical,
92: Elliptical + S0,
93: Elliptical + Spiral,
94: Elliptical + Irregular,
96: S0 + Spiral,
97: Spiral + Spiral,
98: S0 + Irregular,
99: Spiral + Irregular],
(9)~Heliocentric redshift (\kms); (10)~Error in redshift (\kms); 
(11)~Source of redshift: F (FLAIR), T (ANU 2.3m Telescope), N (NED), 
Z (ZCAT, December 2000 version).
\end{minipage}
\end{table*}

\section{Discussion}
\label{sec:discussion}

\begin{figure}
\epsfig{file=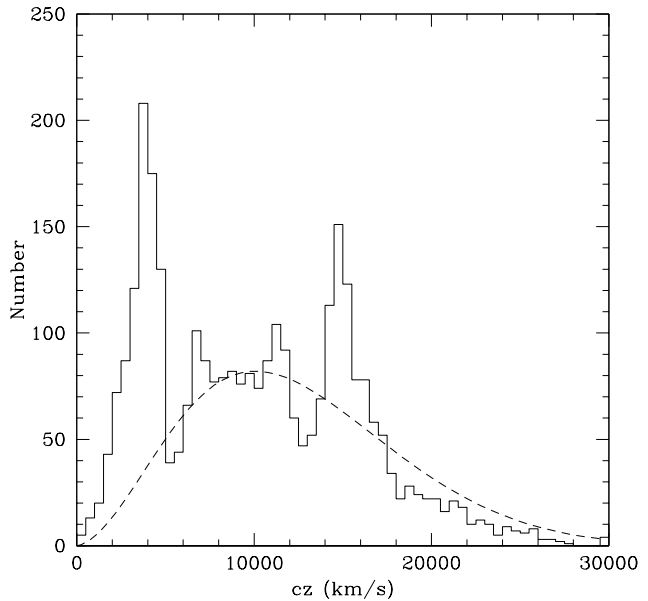,width=\linewidth}
\caption{The redshift distribution for all the galaxies in the FLASH
survey with measured redshifts. The dashed line is the distribution
expected from a homogeneous universe with the same luminosity function
parameters and selection function as the FLASH survey.}
\label{fig:nz}
\end{figure}

The redshift distribution of the FLASH survey is shown in
Figure~\ref{fig:nz}. It extends out to about 30000~\kms\ and has a median
depth of about 10000~\kms. The APM-Stromlo~\cite{loveday:1992} and
Durham/UKST~\cite{ratcliffe:1998} redshift surveys cover factors of 6
and 2 times more sky respectively to a similar depth. However both these
surveys use sparse-sampling strategies (1-in-20 and 1-in-3
respectively), which allow greater sky coverage at the expense of reducing
the resolution of the large-scale structure. The FLASH survey, with a
sampling rate of about 2-in-3, maps a larger number of galaxies in a
smaller volume, thereby sampling the large-scale structure at a higher
resolution and allowing a finer probe of the large-scale structure in
the direction of the \shap\ and Hydra cluster, and better determining
the effects of these structures upon the Local Group motion.

Figure~\ref{fig:nz} also shows the redshift distribution expected for a
homogeneous universe with the same luminosity function parameters (as
derived in a subsequent paper in this series) and selection function as
the FLASH survey. In comparison to this homogeneous redshift
distribution, the most outstanding features in the observed redshift
distribution are the strong peaks due to the many foreground clusters at
$\sim$4000~\kms\ (including the Hydra cluster) and the \shap\ at
$\sim$15000~\kms. 

\begin{figure}
\epsfig{file=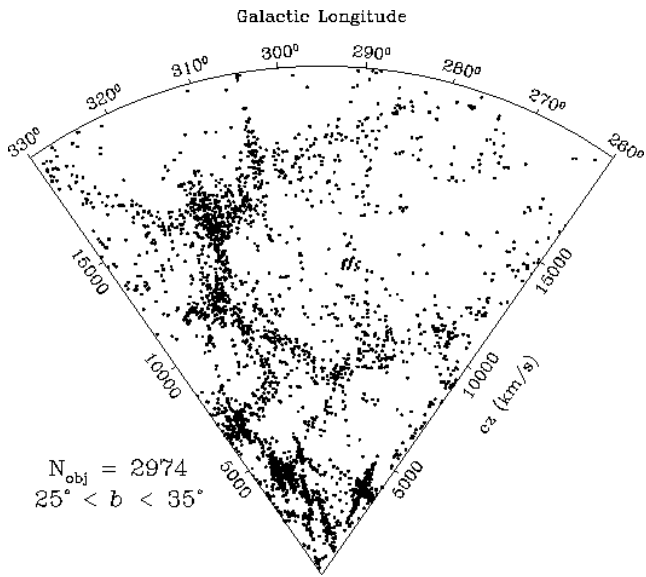,width=\linewidth}
\caption{Redshift wedge plot of all the galaxies in the FLASH survey,
including those with unknown morphologies.}
\label{fig:wedge_all}
\end{figure}

These structures and others are clearly visible in the wedge plot shown
in Figure~\ref{fig:wedge_all}. In this figure, the \shap\ is the large
\fog\ at $l \approx 310\degr$ and $cz \approx 14000$~\kms, while the
Hydra cluster is the sharply defined \fog\ at $l \approx 270\degr$ and
$cz \approx 4000$~\kms. There are many other clusters in the survey
region, evidenced by the tell-tale \fogs\ in Figure~\ref{fig:wedge_all},
most of which are Abell clusters in the direction of the \shap\ (\cf\
Figure~\ref{fig:skyzplot}). Along with the clusters, a number of
wall-like structures can be seen, particularly in the vicinity of the
\shap, where they form a structure similar to the famous `stick-man'
structure associated with the Coma cluster, that was found in the CfA
redshift survey~\cite{cfa:1986}.

Since the FLASH survey has morphological information, it is possible to
examine the properties of each morphological type separately. We choose to
divide the sample into two broad morphological types---early (E and S0) and
late (Sp and Irr)---even though the type information we possess is more
detailed.  The reason for doing so is partly to minimize the effects of
classification errors, but mainly so that there are enough objects in each
grouping to allow meaningful comparisons.

\begin{figure}
\epsfig{file=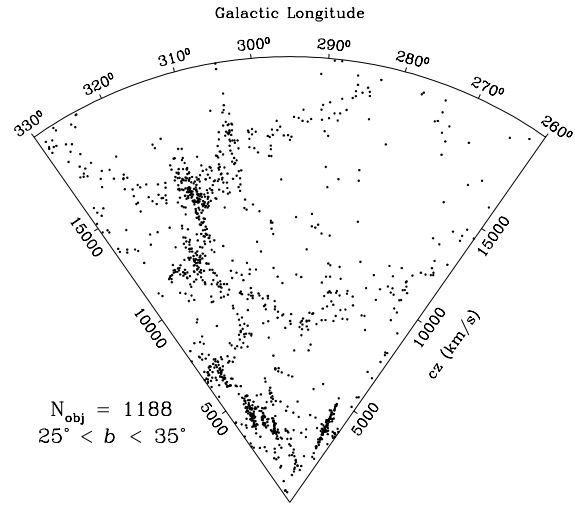,width=\linewidth}
\caption{Redshift wedge plot for the early morphological types in the
FLASH survey.}
\label{fig:wedge_early}
\end{figure}

\begin{figure}
\epsfig{file=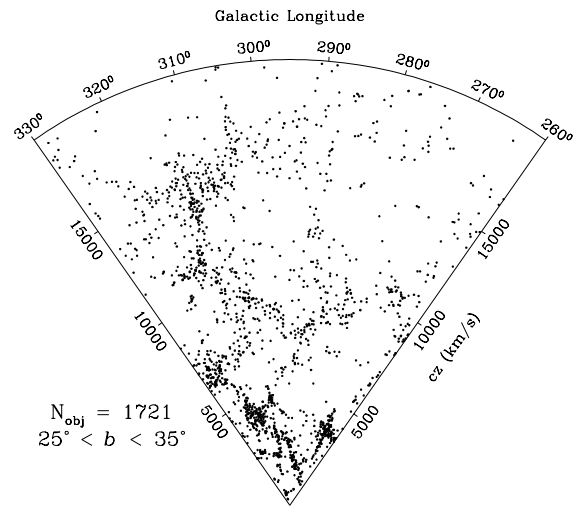,width=\linewidth}
\caption{Redshift wedge plot for the late morphological types in the
FLASH survey.}
\label{fig:wedge_late}
\end{figure}

The redshift wedge plots of the early and late galaxy types are shown in
Figures~\ref{fig:wedge_early} and~\ref{fig:wedge_late} respectively. The
FLASH survey contains redshifts for \nearlyz early-type galaxies, and
\nlatez
late-types. Although similar, the distributions of early and late types
show some interesting differences. Figure~\ref{fig:wedge_early} shows
that the early-type galaxies are found predominantly within cluster
cores, and less frequently within wall-like structures.
The late-type galaxies are found in both clusters and wall-like
structures, as shown in Figure~\ref{fig:wedge_late}, and thus can be
considered as more representative tracers of the large-scale structure.
A good example of this is the bridge of galaxies between $10000 \leq cz
\leq 15000$~\kms along $l = 290\degr$, which is almost entirely populated
with late-type galaxies, and is not visible in
Figure~\ref{fig:wedge_early}.

\section{Conclusion}
\label{sec:conclusion}

The FLASH survey catalogue contains \nflair galaxies brighter than
$\bJ$=16.7 within a $60\degr \times 10\degr$ region bounded by Galactic
latitude $l$=260\degr--330\degr\ and Galactic longitude
$b$=25\degr--35\degr. The galaxies are drawn from the the \hcc\ of
\somak~\shortcite{somak:1990,somak:2002}. The FLASH survey region is richly
structured, and includes the \shap\ and Hydra cluster, as
well as many other Abell clusters, and the \sgp.

The FLAIR-II fibre spectrograph at the UKST was used to obtain redshifts
accurate to $\sim$100~\kms\ for 1896 galaxies. A further 1245 redshifts were
obtained from the NED and ZCAT catalogues, bringing the total number of
galaxies with redshifts in the survey to \nz. The overall redshift
completeness of the sample is 68 per cent. Although the completeness varies
with apparent magnitude, dropping almost to 50 per cent at the faint survey
limit, follow-up observations show that there is no explicit redshift bias.
This magnitude-dependent incompleteness can therefore be corrected in a
straightforward manner.

By selection, the FLASH survey region is not a fair sample of the
universe. The redshift wedge plot for the whole survey region
(Figure~\ref{fig:wedge_all}) reveals a remarkable amount of structure.
The dominant structures are galaxy clusters, shown as \fogs\ in the wedge
plot, with the largest structure being the \shap. Walls and filaments
connect the various clusters, forming in the vicinity of the \shap\ a
structure similar to the `stick-man' structure observed around the Coma
cluster in the CfA survey, but on a much larger scale.

Different structures can be seen in the wedge plots for the early and
late morphological types in the FLASH survey. The most representative
tracer of large-scale structure are the late-type galaxies, which are
associated with both the clusters and the wall-like structures. The
early-type galaxies are less frequently associated with the wall-like
structures, tending instead to be found predominantly in the cores of
rich clusters.

Future papers in this series will examine the galaxy luminosity and
correlation functions, with particular attention to the variations with
morphological type, and reconstruct the peculiar velocity field in this
volume of space in order to determine the contribution of the observed
structures to the peculiar motion of the Local Group.

\section*{Acknowledgements}
We thank the Schmidt support staff and supporting astronomers
who assisted during fibering and observations.
We also thank the UKST unit of the Royal Observatory,
Edinburgh, for supplying us with the relevant survey plate
material, and Mike Irwin and the APM staff for making the scans 
available to us.  
This research has made use of the
NASA/IPAC Extragalactic Database (NED) which is operated by JPL,
Caltech, under contract with NASA.

\bsp

\label{lastpage}

\end{document}